\documentstyle{kluwer} 
\include{psfig}

\newcommand{\sixth}{{\textstyle \frac{1}{6}}}

\newcommand{\onesix}{{\textstyle \frac{1}{6}}}
\newcommand{\onetfour}{{\textstyle \frac{1}{24}}}
\newcommand{\onehtw}{{\textstyle \frac{1}{120}}}

\newcommand{\rn}[1]{(\ref{#1})}
\newcommand{\be}{\begin{equation}}
\newcommand{\ee}{\end{equation}}
\newcommand{\bea}{\begin{eqnarray}}
\newcommand{\eea}{\end{eqnarray}}
\newcommand{\bean}{\begin{eqnarray*}}
\newcommand{\eean}{\end{eqnarray*}}
\newcommand{\half}{{\textstyle \frac{1}{2}}}
\newcommand {\apj} {{\it Astrophys. J.}}
\newcommand {\mnras} {{\it Mon. Not. R. astr. Soc.}}
\begin{document}
\begin{article}
\begin{opening}
\title{STAR CLUSTER SIMULATIONS:~ THE STATE OF THE ART} 
\author{SVERRE J. AARSETH}
\institute{Institute of Astronomy, Madingley Road, Cambridge, England} 
\runningtitle{Star Cluster Simulations}
\runningauthor{Sverre J. Aarseth}
\begin{abstract} 

This paper concentrates on four key tools for performing star cluster
simulations developed during the last decade which are sufficient to
handle all the relevant dynamical aspects. First we discuss briefly the
Hermite integration scheme which is simple to use and highly efficient
for advancing the single particles. The main numerical challenge is in
dealing with weakly and strongly perturbed hard binaries. A new treatment
of the classical Kustaanheimo-Stiefel two-body regularization has proved
to be more accurate for studying binaries than previous algorithms based
on divided differences or Hermite integration. This formulation employs
a Taylor series expansion combined with the Stumpff functions, still with
one force evaluation per step, which gives exact solutions for unperturbed
motion and is at least comparable to the polynomial methods for large
perturbations. Strong interactions between hard binaries and single stars
or other binaries are studied by chain regularization which ensures a
non-biased outcome for chaotic motions. A new semi-analytical stability
criterion for hierarchical systems has been adopted and the long-term
effects on the inner binary are now treated by averaging techniques for
cases of interest. These modifications describe consistent changes of
the orbital variables due to large Kozai cycles and tidal dissipation.
The range of astrophysical processes which can now be considered by N-body
simulations include tidal capture, circularization, mass transfer by
Roche-lobe overflow as well as physical collisions, where the masses and
radii of individual stars are modelled by synthetic stellar evolution.
\end{abstract} 

\end{opening}

\section{Introduction}

The study of self-gravitational $N$-body systems by direct integration
poses many technical challenges which must be addressed.
However, progress during the last decade now enables such problems to be
tackled with confidence.
In this personal review of recent developments, we concentrate on four
main numerical tools which appear to be sufficient for the task in hand.
The corresponding algorithms may be summarized under the following headings:

\medskip
{$\bullet$}~Hermite integration

{$\bullet$}~Two-body regularization

{$\bullet$}~Chain regularization

{$\bullet$}~Hierarchical systems
\medskip

These topics are discussed briefly in the subsequent sections,
together with an outline of current applications.
Given adequate tools, a massive effort is still required in order to develop
an efficient star cluster simulation code but these aspects are beyond the
scope of the present contribution.

\section{Hermite Scheme}

Although the Hermite integration scheme was developed for the
special-purpose HARP computer (Makino 1991), it is also proving highly
effective for standard workstations as well as conventional supercomputers.
Since coding is now somewhat simpler than for the traditional divided
difference formulation (Ahmad and Cohen 1973, Aarseth 1985), it should
be considered the method of choice for direct N-body simulations.
It may also be remarked that Hermite integration is actually more
accurate than divided differences for the same order.

The main idea is again to employ a fourth-order force polynomial but now
the {\it two} first terms are evaluated by explicit summation over all $N$
particles, thereby enabling two corrector terms to be formed.
At first sight it may seem rather expensive to extend the full summation
to the force derivative since this also requires prediction of velocities.
However, simplicity as well as increased accuracy combine to outweigh the
drawback of extra operations, particularly if block-step predictions are
introduced.
We expand a Taylor series for the force $\bf F$ and its first derivative
${\bf F}^{(1)}$ for each particle up to the third derivative about the
reference time $t$ as
\bea
{\bf F} &=& {\bf F}_0 + {\bf F}_0^{(1)} t + 
\half {\bf F}_0^{(2)} t^2
+ \sixth {\bf F}_0^{(3)} t^3 \label{F}, \\
{\bf F}^{(1)} &=& {\bf F}_0^{(1)} + {\bf F}_0^{(2)} t +
\half {\bf F}_0^{(3)} t^2. \label{FD}
\eea
Substituting ${\bf F}_0^{(2)}$ from \rn{FD} into \rn{F} and
simplifying then yields the third derivative corrector
\be
{\bf F}_0^{(3)} = \bigl ( 2 ({\bf F}_0 - {\bf F}) +
({\bf F}_0^{(1)} + {\bf F}^{(1)}) t \bigr ) {6 \over {t^3}}. \label{F3D}
\ee
Similarly, substituting \rn{F3D} into \rn{F} gives the second derivative
corrector
\be
{\bf F}_0^{(2)} = \bigl (-3 ({\bf F}_0 - {\bf F}) -
(2 {\bf F}_0^{(1)} + {\bf F}^{(1)}) t \bigr ) {2 \over {t^2}}. \label{F2D}
\ee

   Using ${\bf F}_0$ and ${\bf F}_0^{(1)}$ evaluated at the beginning of a
time-step, the coordinates and velocities are first predicted to order
${\bf F}^{(1)}$ for all particles.
Following determination of the new ${\bf F}$ and ${\bf F}^{(1)}$ by 
summation over all the contributions, the two higher derivatives are
obtained by \rn{F3D} and \rn{F2D}.
This gives rise to corrector terms for coordinates and velocities
given by
\bea
\Delta {\bf r}_i &=& \onetfour {\bf F}_0^{(2)} \Delta t^4 +
\onehtw {\bf F}_0^{(3)} \Delta t^5, \nonumber \\
\Delta {\bf v}_i &=& \onesix {\bf F}_0^{(2)} \Delta t^3 +
\onetfour {\bf F}_0^{(3)} \Delta t^4.
\eea
Given the high-order derivatives, individual time-steps can now be assigned
in the usual way from some suitable convergence criterion.

The overheads of predicting $N$ coordinates and velocities at each
time-step can be reduced considerably by adopting so-called
hierarchical time-steps (McMillan 1986), where the indicated values are
truncated to be factor 2 commensurate.
The apparent inefficiency of just a few particles sharing the same (small)
step and yet requiring one full prediction is compensated by having a
distribution of discrete levels (typically 16 for $N \simeq 10^4$) such 
that the number of predictions is significantly reduced with respect to
the continuous case (say by factor of 100).
This scheme is particularly suitable for the special-purpose HARP computers
but lends itself equally well to other architectures, including parallel
supercomputers (Spurzem 1998).
Somewhat surprisingly, the workstation code {\tt NBODY6} which is based on
the Ahmad-Cohen (1973) neighbour scheme (Makino and Aarseth 1992, Aarseth 
1994) is in fact slightly faster and more stable than the older {\tt NBODY5}
code for $N = 1000$ single particles and the same number of steps.

\section{Two-Body Regularization}

The early 1970's saw the introduction of the Kustaanheimo-Stiefel (1965)
regularization for treating close encounters and hard binaries in $N$-body
simulations (Bettis and Szebehely 1972, Aarseth 1972) and
the elegant KS method has proved to be very resilient.
However, even a regularized two-body solution is subject to small but
systematic errors when studied over long times.
In order to avoid this undesirable feature, the concept of energy
stabilization has been tried for weak perturbations (Aarseth 1985).
Although this procedure ensures that the orbit is constrained to have the
correct energy arising from the perturbation, the corresponding angular
momentum is no longer conserved so well.

The subsequent exploitation of adiabatic invariance (Mikkola and Aarseth
1996) by the so-called slow-down principle tends to alleviate this
imperfection since now one KS orbit may represent a number of physical 
periods by augmenting the perturbation itself and neglecting short-period
effects.
As for the earlier claim that a time-symmetric KS method would be superior
(Funato et al. 1996), it now appears that the requirement of variable
time-steps for perturbed orbits cannot be accommodated (Kokubo et al. 1998).
So far there is no evidence that the resulting eccentricities of cluster
binaries studied by the stabilization scheme cannot be trusted, especially
bearing in mind that the long-term evolution of most binaries is
predominantly subject to discrete changes of a random nature.
The case of long-lived hierarchical systems deserves special consideration,
however, but here additional effects should also be considered, as
discussed in a subsequent section.

An alternative KS regularization scheme has been presented recently
(Mikkola and Aarseth 1998) which achieves a high accuracy without extra
cost.
This new approach is based on the idea of a truncated Taylor series, where
additional correction terms represent the neglected higher orders and
which yields exact solutions in the unperturbed case.
The new algorithm is again of Hermite type and will be outlined in the
following.

First, coordinates and velocities of the perturbers are predicted in the
usual way (i.e. to first order), whereas the regularized coordinates
and velocities (${\bf U, U}'$) are predicted to highest order.
Here ${\bf U}^{(4)}, {\bf U}^{(5)}$ include the modified Stumpff (1962)
functions
\be
 \tilde c_n(z) = n! \sum_{k=0}^\infty \frac{(-z)^k}{(n+2k)!},
\ee
where the argument is related to the time-step by
$z = -\half h \Delta \tau^2$ and $h$ is the specific binding energy.
These coefficients only deviate slightly from unity and a twelfth-order
expansion (re-evaluated every step) appears sufficient.
After transforming the physical coordinates and velocities to global values,
the predictor cycle is completed by evaluating the perturbing acceleration
$\bf F$ as well as its explicit derivative $\bf \dot F$.

Because of the insufficient accuracy of the predicted deviation from
unperturbed motion at the end of a step, the corrector cycle employs an
iteration.
Setting $\Omega = -\half h$, the basic equation of motion takes the
familiar form
\be
{\bf U}^{(2)} = -\Omega {\bf U} + \half r {\cal L}^T {\bf F},
\ee
where $\cal L ({\bf U})$ is a $4 \times 3$ linear matrix and
$r = {\bf U} \cdot {\bf U}$ is the separation.
We express the new KS acceleration and its derivative
(where ${\bf F}' = r {\bf \dot F}$) at the start of a step as
\bea
{\bf U}_0^{(2)} &=& -\Omega_0 {\bf U}_0 + \bf f_0^{(2)}, \label{U2} \\
{\bf U}_0^{(3)} &=& -\Omega_0 {\bf U}_0' + \bf f_0^{(3)}, \label{U3}
\eea
where ${\bf f}_0^{(2)} = \half r {\bf Q}$, with $\bf Q = {\cal L}^T{\bf F}$,
is the perturbed force contribution evaluated after the {\it previous}
predictor cycle.

The two next Taylor series terms are constructed from the Hermite scheme.
Using the current value of $h$ (and $\Omega$), predicted to fourth order,
we form the new perturbative functions at the end of the step
\bea
\bf f^{(2)} &=& (\Omega_0 - \Omega) {\bf U} +\half r {\bf Q}, \label{F2} \\
\bf f^{(3)} &=& (\Omega_0 - \Omega) {\bf U}' - \Omega' {\bf U} +
 \half r' {\bf Q} +\half r {\bf Q}', \label{F3}
\eea
from which the corrector derivatives $\bf f_0^{(4)},\bf f_0^{(5)}$ are
recovered by the Hermite rule (Makino 1991).

The expressions for ${\bf U}_0^{(4)}$ and ${\bf U}_0^{(5)}$ are readily
formed in analogy with Eqs.~\rn{U2} and \rn{U3} which yield
\bea
{\bf U}_0^{(4)} &=& -\Omega_0 {\bf U}_0^{(2)} + \bf f_0^{(4)}, \label{U4} \\
{\bf U}_0^{(5)} &=& -\Omega_0 {\bf U}_0^{(3)} + \bf f_0^{(5)}. \label{U5}
\eea
From Eqs.~\rn{U2} - \rn{U5},
the provisional solution for $\bf U, U'$ is then obtained by the general
expression (cf. Mikkola and Aarseth 1998), which contains the Stumpff
functions.
The treatment of the energy remains the same as for standard Hermite
based on $\Omega ' = -{\bf U}' \cdot {\bf Q}$ and the physical time is
obtained from integrating $t' = {\bf U} \cdot {\bf U}$ which also involves
Stumpff functions.
Substituting for ${\bf U}^{(2)}$, we write the second derivative as
\be
\Omega^{(2)} = \Omega_0 {\bf U} \cdot {\bf Q} - {\bf f}^{(2)} \cdot {\bf Q}
 - {\bf U}' \cdot {\bf Q}'.
\ee
The two corrector terms constructed from $\Omega'$ and $\Omega^{(2)}$
are added to the predicted value
{\it without} any Stumpff functions to yield an improved solution for
$\Omega$ at the start of the next iteration or at the end point.

Subsequent iterations repeat the procedure above, starting from
Eq.~\rn{F2} without re-evaluating the physical perturbation and
its derivative.
Thus the new values of Eqs.~\rn{F2} and \rn{F3} are based on the
improved solution for $\bf U, U'$ and $r, r'$, as well as the new $\Omega$.
In the present treatment, one iteration yields a significant improvement
for modest perturbations and experience so far indicates that this may
also be sufficient for strong interactions because of the shortening of
the stepsize $\Delta \tau$ (cf. Aarseth 1994).

The corrector cycle ends by specifying new derivatives for use in the
next prediction, as well as saving the perturbative derivatives
\rn{F2} and \rn{F3} required for the Hermite scheme.
This is completed by re-initializing Eqs.~\rn{U2} and \rn{U3}
at the end point, substituting ${\bf f}^{(2)}, {\bf f}^{(3)}$
as well as the iterated values of $\Omega'$ and $\Omega^{(2)}$.
It is advantageous to employ the corrected values of $r$ and $r'$ for
this purpose; the re-evaluation of ${\bf f}^{(2)}$ and ${\bf f}^{(3)}$
is fast and also benefits the final quantities
${\bf U}^{(2)}$ and ${\bf U}^{(3)}$ to be used in the next prediction.
A more accurate expression of the fourth KS derivative at the
{\it end} of the interval is obtained by including the next order by
\be
{\bf U}^{(4)} = {\bf U}_0^{(4)} + {\bf U}_0^{(5)} \Delta \tau,
\ee
and similarly for the {\it third} derivative of the energy,
\be
\Omega^{(3)} = \Omega_0^{(3)} + \Omega^{(4)} \Delta \tau.
\ee

The above scheme has been implemented in the state of the art codes
{\tt NBODY4} and {\tt NBODY6} and has proved itself in large-scale
simulations.
Accuracy tests obtained by a toy code shows that high accuracy can be
obtained with 30 steps per orbit for relatively weak perturbations, which
is about half that required by the old stabilization scheme.
A significant part of this gain is due to the modifications by the
Stumpff coefficients, although the basic Taylor series (or Encke-type)
formulation is also considerably more accurate than the standard method.
The number of operations for a typical step is not much larger in the
new method, including the overhead for the Stumpff functions and one
iteration in the corrector.
Hence the computational effort is less for typical calculations,
although this depends on the actual number of perturbers.
Finally, we remark that the Stumpff method also includes the slow-down
scheme in exactly the same way as before.

\section{Chain Regularization}

The concept of chain regularization is simple, yet the mathematical
formulation is quite technical and this has acted as an impediment to
wider usage.
However, it enables new types of problems to be studied and is therefore
worth the extra effort.
The basic idea is a generalization of three-body regularization
(Aarseth and Zare 1974) which treats two perturbed KS solutions with
respect to a common reference body, where each two-body solution is
described by regular equations.
Thus an extension to four participating bodies merely introduces one more
perturbed KS solution, although the formalism is somewhat different
(Mikkola and Aarseth 1990).
Once the step from three to four particles has been mastered, the general
case becomes feasible (Mikkola and Aarseth 1993).

The essential feature of chain regularization is that dominant interactions
along the chain itself are treated as perturbed KS solutions and all the
other attractions are included as perturbations.
Hence it becomes imperative to select the chain vectors in such a manner
as to minimize the perturbations.
Since we are dealing with dynamical interactions, the chain vectors need
to be redrawn in response to changing configurations.
Fortuitously, all the relevant decision-making constitute a minor
overhead here since the integration is carried out by the high-order
Bulirsch-Stoer (1966) scheme and a certain elasticity is tolerated as
regards switching to more favourable chain vectors.

The equations of motion are derived from a regularized Hamiltonian of
the form
\be
 \Gamma^{\star} = g (H - E),
\ee
where $H$ is expressed in terms of the coordinates and momenta and $E$
is the internal system energy.
Here the function $g$ is given by the corresponding time transformation 
\be
 dt = g d \tau
\ee
and choosing the inverse Lagrangian energy ($L = T + \Phi $) ensures
regular solutions for any chain separation $R_k$.

The treatment begins by selecting a compact subsystem of three or four
particles; i.e. so-called $B + S$ or $B + B$ type.
External perturbers are chosen in analogy with the KS implementation and
the internal integration includes any perturbation effect which also tends
to change the total energy according to its separate equation of motion.
At the same time, the c.m. motions are advanced by the standard Hermite
scheme with due attention to the slightly modified form of the
corresponding acceleration which requires a differential correction.

The analogy with KS does not hold in one important respect since the
chain membership may change before termination occurs.
Thus an initial subsystem of four members may lose one member due to
ejection, or an approaching perturber - a single particle or binary -
may be added.
Alternatively, the membership may also change through physical collision.
All the relevant corrections and re-initializations are performed
{\it in situ}.
Hence the use of chain variables is also highly beneficial for the
evaluation of nearly singular quantities.
Chain termination usually occurs when a binary becomes well separated
from one or two other members in which case the binary is accepted for
KS treatment, whereas the remaining membership is initialized by the
Hermite scheme or even as a second KS system.
The actual decision-making also takes into account the cluster environment
and is therefore quite involved.

Cluster simulations of primordial binaries frequently involve interactions
of two binaries where the size of one is much less than the other.
In such cases even the powerful chain method becomes prohibitive because
the shortest period is a small fraction of the local crossing time.
Fortunately the principle of slow-down applied to weakly perturbed KS
binaries can also be employed here (Mikkola and Aarseth 1996).
This permits a consistent study of binaries with arbitrarily short periods
which would otherwise have to be treated as inert systems.
The implementation itself differs from the KS case since here we adjust
the slow-down factor continuously according to the maximum apocentre
perturbation exerted by the other chain members, rather than choosing an
appropriate discrete level (factor of 2) at each apocentre passage.

Since the strong interactions studied by the chain method are usually of
short duration, the simulation code only allows one such case to be
considered at a time for technical reasons.
However, there is provision for studying one triple as well as one
quadruple system by {\it unperturbed} three-body (Aarseth and Zare 1974)
and chain (Mikkola and Aarseth 1990) regularization.
Given a few hundred critical events in a typical cluster simulation, the
latter procedures are usually not needed but this may change with the
addition of more primordial binaries.

\section{Hierarchical Systems}

The Solar neighbourhood contains many examples of multiple systems where
the inner component of a binary is itself a binary, and levels of higher
multiplicity also exist.
Likewise, hard binaries in star clusters may acquire an outer component
with sufficiently small eccentricity to be stable over many orbits.
Hierarchical triples may be formed by the classical three-body capture
mechanism in which the binary itself acts mainly as a point-mass.
However, in clusters with significant binary populations such systems are
more likely to form in strong interactions between two binaries since this
involves two-body encounters.
The second formation process was already identified in scattering
experiments with colliding binaries which yielded a high percentage of
positive outcomes (Mikkola 1983).
Thus one way for such triples to become stable requires the impact
parameter to exceed some critical value and yet be sufficiently small for
the weakest binary to be disrupted, but other processes are also favoured,
including exchange.

Given a newly formed hierarchical triple, the question of long-term
stability naturally arises.
Depending on the period ratio, the direct calculation of a perturbed
inner binary can be quite time-consuming even with KS regularization.
However, since the corresponding semi-major axis may not be subject to
any secular effects it becomes possible to adopt the centre-of-mass
approximation and thereby only neglect cyclical changes of the eccentricity.
Various empirical criteria have been obtained by fitting the results of
systematic three-body calculations for a restricted set of parameters
(Harrington 1977, Eggleton and Kiseleva 1995). 
Based on these results, the so-called merger procedure has been employed
for some time (Aarseth 1985).
Thus provided the stability condition is satisfied, the inner binary is
replaced by its combined mass to facilitate KS treatment of the
{\it outer} orbit.

A more rigorous approach based on correspondence with the chaos boundary in
the binary-tides problem (Mardling 1995) has yielded a semi-analytical
stability criterion
which holds for quite large mass ratios and arbitrary outer eccentricities
(Mardling and Aarseth 1998).
Here the critical outer pericentre distance is given in terms of the 
inner semi-major axis, $a_{in}$, by
\be
{R_p^{out}}
= C\left[(1+q_{out})\frac{(1+e_{out})}{(1-e_{out})^{1/2}}\right]^{2/5}a_{in}
\label{Stab},
\ee
where $q_{out} = m_3/(m_1 + m_2)$ is the outer mass ratio, $e_{out}$ is
the corresponding eccentricity and $C \simeq 2.8$ is determined empirically.
This criterion is only valid for coplanar prograde motion and still ignores
a weak dependence on the inner eccentricity.
However, the general case of inclined orbits exhibit increased stability
so that Eq.~\rn{Stab} represents an upper limit.
Further tests suggests an inclination correction factor
$f = 1 - 0.3 i /180$ (with $i$ in degrees) which has been adopted in
practical simulations; this is also in qualitative agreement with the
original stability condition for retrograde orbits (Harrington 1972).
The merger treatment is only allowed while the pericentre condition is
satisfied, after which the inner binary is re-initialized.

A further refinement is included when the outer component itself is a
binary.
In the case of a $B + B$ configuration, the smallest binary plays the role
of the outer body in a triple.
Since the corresponding chaos boundary is not very sensitive to a second
extended object (Mardling 1991), we adopt an additional correction factor
$f_1 = f + 0.1 min (a_{in} /a_2, a_2 /a_{in})$, with $a_2$ representing the
second semi-major axis.
We also mention here that even double hierarchies may be formed, where a
system of type $B + S$ or $B + B$ itself acquires an outer bound component.
Such configurations do occur occasionally and procedures have therefore
been developed for their special treatment.

The criterion \rn{Stab} above is concerned with long-term stability and
hence the absence of escape.
However, it is also of interest to consider the possibility of exchange
between the outer component and one member of the inner binary.
According to classical developments (Zare 1977, Szebehely and Zare 1977),
the critical value for exchange in a coplanar prograde triple is given by
\be
(c^2 E)_{crit} = -\frac{G^2 f^2(\rho) g(\rho)} {2 (m_1 + m_2 + m_3)},
\ee
where $\bf c$ is the angular momentum and the functions $f(\rho), g(\rho)$
are expressed in terms of the masses, with $\rho$ determined by iteration
from a fifth-order algebraic equation for the collinear equilibrium points.
Numerical tests show that the chaos boundary given by Eq.~\rn{Stab} lies
above the exchange boundary when the masses are comparable and the latter
only begins to overlap above $q_{out} \simeq 5$.
Application of the exchange criterion is therefore less useful in practical
calculations.
We also note that once an exchange occurs the final evolution will
inevitably lead to escape.

The long-term evolution of a hierarchical triple is characterized by cyclic
oscillations of the inner eccentricity where the amplitude depends on the
inclination.
The so-called Kozai effect (Kozai 1962) has received much attention
recently in connection with external planetary systems but there is also
an early example from N-body simulations (van Albada 1968) which points
to the relevance for star clusters.
Various analytical tools have been employed in order to model this process
in some detail, including tidal dissipation for high eccentricities
(Mardling and Aarseth 1999).
Among the useful quantities which can be calculated theoretically
(Heggie 1995) are the time-scale for a complete oscillation, $T_K$, as well
as its maximum value, $e_{max}$.

Since the time-scale for the Kozai cycle is usually much greater than the
Kepler period, the merger scheme for hierarchical triples lends itself
particularly well to a semi-analytical treatment.
At present only systems with $e_{max} > 0.8$ are considered since smaller
amplitudes are less likely to result in tidal activity.
We have used a double averaging procedure (Eggleton 1997, Mardling and
Eggleton 1998) to calculate the evolution of such systems in terms of the
inner Runge-Lenz vector and angular momentum vector.
Thus some examples show that inclinations near $90^{\circ}$ may induce
tidal circularization even if oblateness effects are included.
Clearly further developments of this experimental approach is needed in 
order to improve the modelling of these complicated processes.

\section{Astrophysical Applications}

The realistic simulation of star cluster dynamics requires a variety of
astrophysical processes to be considered.
In particular, the implementation of consistent stellar evolution enables
the study of mass loss and finite-size effects.
This is an ongoing project which has been outlined elsewhere
(Aarseth 1996) and now employs an improved description of Roche mass
transfer and physical collisions (Tout et al. 1997).
Particular emphasis has been devoted to the modelling of chaotic motions
and tidal circularization which form a link between an initial binary
distribution and the Roche stage (Mardling and Aarseth 1999).
In particular, it is found that very high eccentricities ($e > 0.999$)
are produced in stable hierarchies or by exchange and these in turn
lead to orbital shrinkage by tidal dissipation.
Primordial binaries also leave an imprint in the form of high-velocity
escapers.
At the same time, more general cluster simulations have yielded much
insight into dynamical evolution (McMillan et al. 1992,
Aarseth and Heggie 1998, Portegies Zwart et al. 1998).

\begin{figure}[t]
\centerline{\psfig{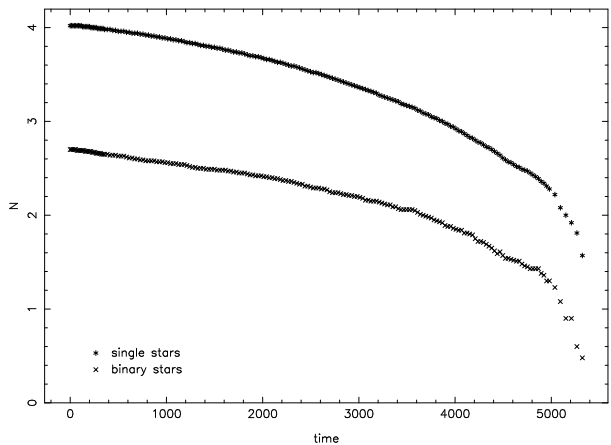}}
\caption{Logarithmic plot of single stars and binaries as functions of time
in $Myr$.}
\end{figure}

The modelling of synthetic stellar evolution is based on fast look-up
tables for the radius, luminosity and type as a function of the initial
mass and age (cf. Tout et al. 1997).
Instantaneous mass loss due to stellar winds or supernovae explosions are
adopted for the advanced stages.
An energy-conserving integration scheme is preserved by including relevant
corrections and re-initializations.
The standard open cluster model includes $10^4$ single stars with $5 \%$
primordial hard binaries.
Once the most massive single stars have evolved, the binaries dominate
the mass segregation and increase their central abundance significantly
with increased disruption probability.
Even so, the original binary population is not depleted preferentially
such that there is always an energy source which prevents core collapse.
This behaviour is illustrated well in the figure which displays the
bound membership.

In conclusion, the algorithms presented above have proved highly efficient
for star cluster simulations.
Hopefully these numerical tools will also play a part in future efforts
involving more powerful hardware.

\label{lastpage}
\end{article}
\end{document}